# 3D Reconstruction of Coronary Arteries and Atherosclerotic Plaques based on Computed Tomography Angiography Images

Vassiliki I. Kigka[1,2], George Rigas[1,2], Antonis Sakellarios[1,2], Siogkas Panagiotis[1,2], Ioannis O. Andrikos[1,2], Themis P. Exarchos[1,2] Dimitra Loggitsi[3], Constantinos D. Anagnostopoulos[4], Lampros K. Michalis[5], Danilo Neglia[6], Gualtriero Pelosi[7], Oberdan Parodi[7], Dimitrios I. Fotiadis[1,2,*]

[1]Unit of Medical Technology and Intelligent Information Systems, Department of Materials Science and Engineering, University of Ioannina, GR 45110, Ioannina, Greece.

[2]Institute of Molecular Biology and Biotechnology, Dept. of Biomedical Research Institute – FORTH, University Campus of Ioannina, GR 45110, Ioannina, Greece.

[3]CT/MRI Department, Hygeia & Mitera Hospitals, Athens, Greece.

[4]PET-CT Department & preclinical Imaging Unit Center for Experimental Surgery, Clinical & Translational Research Biomedical Research Foundation Academy of Athens.

[5]Department of Cardiology, Medical School, University of Ioannina, GR 45110, Ioannina, Greece.

[6]Fondazione Toscana Gabriele Monasterio, Pisa, IT, 56126 Pisa.

[7]Institute of Clinical Physiology, National Research Council, Pisa IT 56124, Italy.

[*]Corresponding author

Unit of Medical Technology and Intelligent Information Systems,

Dept. of Materials Science and Engineering, University of Ioannina,

PO Box 1186, GR 451 10 Ioannina, GREECE

tel.: +30 26510 09006, fax: +30 26510 08889

e-mail: fotiadis@cc.uoi.gr

**Abstract**

The purpose of this study is to present a new semi-automated methodology for three-dimensional (3D) reconstruction of coronary arteries and their plaque morphology using Computed Tomography Angiography (CTA) images. The methodology is summarized in seven stages: pre-processing of the acquired CTA images, extraction of the vessel tree centerline, estimation of a weight function for lumen, outer wall and calcified plaque, lumen segmentation, outer wall segmentation, plaque detection, and finally 3D surfaces construction. The methodology was evaluated using both expert's manual annotations and estimations of a recently presented Intravascular Ultrasound (IVUS) reconstruction method. As far as the manual annotation validation process is concerned, the mean value of the comparison metrics for the 3D segmentation were 0.749 and 1.746 for the Dice coefficient and Hausdorff distance, respectively. On the other hand, the correlation coefficients for the degree of stenosis 1, the degree of stenosis 2, the plaque burden, the minimal lumen area and the minimal lumen diameter, when comparing the derived from the proposed methodology 3D models with the IVUS reconstructed models, were 0.79, 0.77, 0.75, 0.85, 0.81, respectively. The proposed methodology is an innovative approach for reconstruction of coronary arteries, since it provides 3D models of the lumen, the outer wall and the CP plaques, using the minimal user interaction. Its first implementation demonstrated that it provides an accurate reconstruction of coronary arteries and thus, it may have a wide clinical applicability.

**1. Introduction**

Coronary Artery Disease (CAD), in which atheromatic plaques build up inside the coronary arteries, is the most common type of heart disease and is considered as the leading cause of morbidity and mortality world wide [1]. Non-invasive cardiovascular imaging and particularly Computed Tomography Angiography (CTA) has experienced a remarkable progress in the last years [2, 3]. CTA has been gaining widespread acceptance in clinical practice for the investigation of suspected CAD, since it is able to visualize the coronary arteries and their anatomy and allows the interpreter to evaluate the presence, extent and type (calcified (CP) or non-calcified (NCP)) of atherosclerotic plaques [4], without the invasive catheterization procedure.

Different studies have indicated that CTA modality is able to analyze accurately the coronary artery remodeling and provides not only the detection and quantification of the atherosclerotic plaque [5, 6], but also the classification of its composition [7]. In addition to its high accuracy, CTA provides robust prognostic information in patients with suspected CAD and allows the risk stratification as well, when CAD is present [8] while it can be used for prediction of plaque growth based on computational modelling [9, 10]. Existing studies have demonstrated that CTA derived measures, such as the number of vessels with significant stenosis, the luminal stenosis, the stenosis location, the plaque burden and the composition of the plaque contribute to diagnostic and prognostic abilities of coronary CTA [11, 12].

An accurate 3D model of coronary arteries, except for visualizing the vessel geometry and its plaque distribution, allows also the blood flow simulation and the investigation of the role of biomechanical factors, such as static pressure, wall shear stress, blood viscosity on the localization and progression of atherosclerosis [13]. In addition to this, a 3D coronary imaging has the potential to provide a more comprehensive evaluation of the risk for CAD progression [14].

In the literature, different studies have been presented to determine the accuracy of 3D artery reconstruction and the assessment of plaque using CTA. Voros *et al*. [15] presented a study for the evaluation of 3D quantitative measurements of coronary plaque by CTA using Intravascuar

Ultrasound (IVUS). Another similar approach was introduced by Graaf *et al*. [16], who studied the correlation between the metrics derived by CTA automatic software (QAngio CT 1.1, Medis medical imaging systems) and those provided by Virtual Histology IVUS (VH-IVUS), which was defined as the gold standard. Arbab *et al*. [17] performed a study for the quantification of coronary arterial stenosis using CTA and demonstrated that CTA in comparison with the conventional angiography, is able to identify non-invasively patients with CAD. Athanasiou *et al*. [18] presented a semi-automated methodology for 3D reconstruction of arteries and their plaque morphology using CTA images and compared their approach using IVUS findings.

In this work, we present a new semi-automated methodology for 3D coronary artery reconstruction and plaque detection using CTA modality. In order to investigate the accuracy of our approach, we implemented two different validation approaches, using both expert's annotations and estimation of an IVUS reconstruction methodology. The comparison results indicated good agreement. ==The main innovative aspect of the presented methodology is its ability to reconstruct both the lumen, the outer wall and the CP plaques with the minimal user interaction==. Furthermore, the proposed methodology incorporates a centerline extraction, using a minimum cost path approach. Thus, a successful and accurate centerline detection is guaranteed and the subsequent step of lumen segmentation is improved. In addition to this, it indicates a user friendly applicability, since the main user interaction is the detection of the start and the end point of each branch.

## 2. Materials and Methods

The proposed methodology includes 7 stages. In the first stage, the acquired CTA images are pre-processed to detect vessel silhouette. In the second stage, a centerline extraction approach of the vessel is applied. In the third stage, a weight function for the lumen, the outer wall and the CP plaques is estimated. In the fourth stage, an extension of active contour models for lumen segmentation is implemented. In the fifth and sixth stage, similarly to the previous stage, a level set methodology for outer wall segmentation and plaque segmentation, respectively, is applied. Finally, in the last stage the 3D surfaces for the lumen, the outer wall

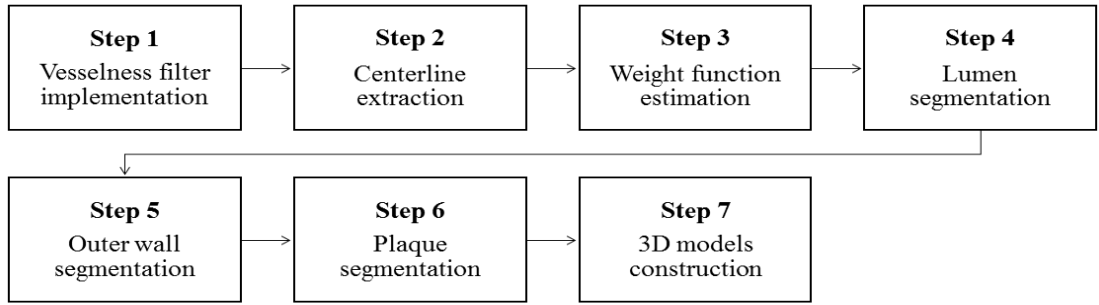

**Figure 1**. The proposed methodology steps starting with the preprocessing implementation and resulting in a lumen, outer wall and plaques 3D model.

and the CP plaques are constructed. In Fig. 1, the stages of the proposed methodology are shown.

**2.1 Preprocessing**

The image preprocessing step is applied in the axial DICOM acquired images to remove irrelevant details of the CTA images. A vessel enhancement filter, the Frangi Vesselness filter [19] is implemented to identify tubular structures and limit the region of interest (ROI) to vessel candidate regions. In Fig. 2, an example of the implementation of the Vesselness filter is shown.

**2.2 Centerline Extraction**

The centerline is mainly required for creating an initial vessel mask for the vessel segmentation algorithm. However, the centerline extraction stage still remains a challenging task, since the size of the vessels is small and several reconstruction artifacts are observed. In the proposed methodology, a minimum cost path approach is implemented for the centerline extraction, based on Metzt *et al.* [20] approach.

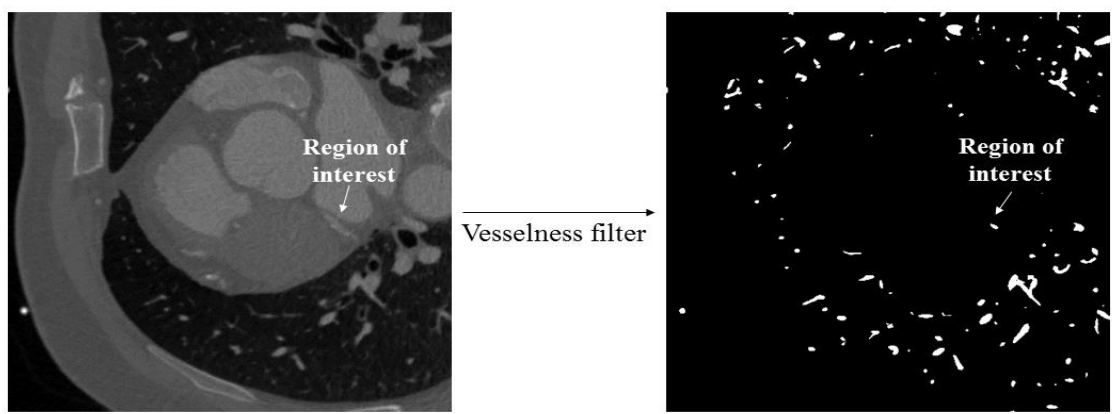

**Figure 2**. An example of the implementation of the Vesselness filter in a CTA image.

The proposed centerline extraction methodology is quite simple and, therefore easy to implement, since the main requirement is the starting point and the ending point of the vessel to extract the corresponding centerline. The cost function, which is considered for the minimum cost path approach is a combination of the lumen and vessel weight.

Firstly, we extract the image weight based on the vesselness measure ($w_{vessel}$) [19]. Subsequently, we compute the value of the top 50% of the image intensities, which are larger than 100 Hounsfield Units (HU), considering only the parts of the image, where the $w_{vessel}$ measure is larger than 0. This computed value $ml$ is very significant, since it is used for the extraction of the lumen weight. More specifically, the lumen weight is extracted by using a generalized bell-shaped membership function and it is defined as:

$$w_{lumen} = 0.9 \cdot \frac{1}{1+\left|\frac{x-c}{a}\right|^{2b}} + 0.1, \qquad (1)$$

where $a = 0.02$, $b$ is the minimum value between $ml - l_{thres}$ and 500, and $c$ is the value of $ml + cp_{thres}$. Heuristically, and making several experiments, the threshold of the lumen ($l_{thres}$) and the calcified plaques ($cp_{thres}$) was defined 80 HU and 400 HU, respectively. More details can be found in the Appendix.

The considered cost function $V$ for the minimum path approach is a combination of the vessel and the lumen weight and is defined by:

$$V = w_{vessel} \cdot w_{lumen}. \qquad (2)$$

In order to calculate the shortest distance from a list of points to all other pixels in an image volume, a Multistencil Fast Marching Method (MSFM) is implemented based on the approach described in [21]. An example of the above procedure is depicted in Fig. 3.

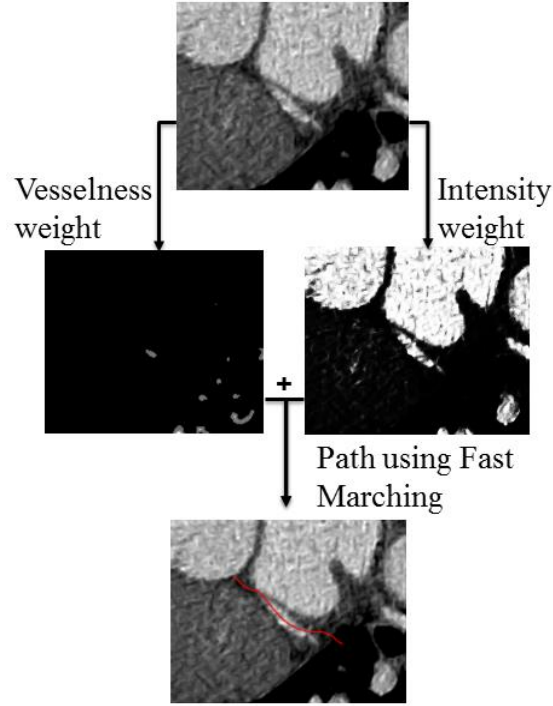

**Figure 3**. Example of a successfully extracted coronary artery centerline using the vesselness/intensity cost function.

**2.3 Estimation of Weight Function for lumen, outer wall and calcified plaque**

Similar to the previous step, in this stage three different membership functions for the lumen, the outer wall and the CP plaques are computed, aiming to compensate different protocols for discriminating the lumen, the outer wall and the calcified plaque. These membership functions are all adapted to the mean vessel intensity across the centerline, assuming that this corresponds to the mean lumen intensity. More specifically, the mean lumen intensity $\bar{I}_{lumen}$ is calculated, taking into consideration only the pixels of the image, whose intensities are higher than 100 HU and their Euclidean distance from the extracted centerline is less than 5.

For the lumen a generalized bell-shaped membership function is used, whereas for the outer wall and the calcified plaque two different sigmoidal membership functions are implemented, as it is shown in Fig. 4. The generalized bell-shaped membership function is defined as:

$$g^{bell}(x;a,b,c) = \frac{1}{1+\left|\frac{x-c}{a}\right|^{2b}}, \qquad (3)$$

whereas the sigmoidal membership function is defined as:

$$g^{sigm}(x;a,b) = \frac{1}{1+e^{-a(x-b)}}, \quad (4)$$

where $x$ is the image and $a,b,c$ are the defined parameters.

For the lumen, the membership function is given by:

$$f_{lumen} = (1-\varepsilon) \cdot g^{bell}(x;a_{lumen},b_{lumen},c_{lumen}) + \varepsilon, \quad (5)$$

where $a_{lumen} = 0.02$, $b_{lumen} = \min\left(\left[\max\left(\left[\bar{I}_{lumen} - l_{thres} \quad 150\right]\right) \quad 500\right]\right) - 0.01$ and $c_{lumen} = \bar{I}_{lumen} + cp_{thres}$.

The membership function for the outer wall and the plaques is in both cases a sigmoidal function and is given by:

$$f_{outer/plaque} = (1-\varepsilon) \cdot g^{sigm}(x;a_{outer/plaque},b_{outer/plaque}) + \varepsilon, \quad (6)$$

where $a_{outer} = 0.02$, $a_{plaque} = 0.05$, $b_{outer} = \min\left(200, \max\left(\left[\bar{I}_{lumen} - l_{thres} - ncp_{thres} \quad 100\right]\right)\right)$ and $b_{plaques} = \bar{I}_{lumen} + cp_{thres}$. The threshold value for the lumen ($l_{thres}$) and the calcified plaques ($cp_{thres}$) as previously stated are 80 HU and 400 HU, respectively. The intensity threshold for non calcified plaques ($ncp_{thres}$) is defined by the value of 50 HU. Similarly to the $l_{thres}$ and the $cp_{thres}$, the value of the $ncp_{thres}$ is defined heuristically and based on the current literature [22, 23]. More details can be found in the Appendix. The parameter $\varepsilon$ is a weight of the membership functions and in all cases it is defined by the value of 0.05. In Table 1, we summarize the different values of the parameters of the membership functions for each component. It is obvious that these parameters values are approximated and there is a need to adapt them into the acquired CTA image intensity.

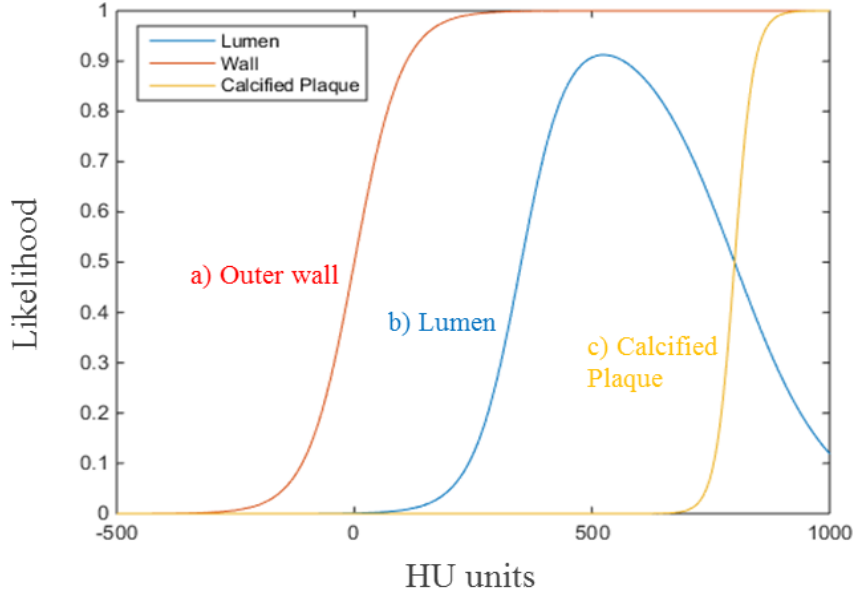

**Figure 4**. Membership functions distributions over HU units for a) outer wall, b) lumen, c) calcified plaques based on HU units.

**Table 1**. A summary of the parameters of the membership functions for the lumen, the outer wall and the CP plaques.

| Parameters | $a$ | $b$ | $c$ |
|---|---|---|---|
| **Lumen** | $a_{lumen} = 0.02$ | $b_{lumen} = \min\left(\left[\max\left(\left[\bar{I}_{lumen} - l_{thres} \quad 150\right]\right) \quad 500\right]\right) - 0.01$ | $c_{lumen} = \bar{I}_{lumen} + cp_{thres}$ |
| **Outer wall** | $a_{outer} = 0.02$ | $b_{outer} = \min\left(200, \max\left(\left[\bar{I}_{lumen} - l_{thres} - ncp_{thres} \quad 100\right]\right)\right)$ | - |
| **CP plaques** | $a_{plaque} = 0.05$ | $b_{plaques} = \bar{I}_{lumen} + cp_{thres}$ | - |

**2.3 Lumen Segmentation**

In this step, an extension of the active contour models [24] is implemented for lumen segmentation. This approach is based on regional measures and does not depend on any edge definition. In other words, the boundaries of the detected objects are not necessarily defined by the gradient. The main improvement of our lumen level set segmentation approach is that it incorporates a prior shape [25], aiming to segment an object whose shape is similar to the given prior shape which is independent of translation, scaling and rotation, from a background where there are several objects. For lumen segmentation, the prior shape is a tubular mask across centerline with a small radius.

*2.3.1   Update of lumen intensities*

Our purpose is to implement the membership function in the part of the CTA image, which is near to the extracted centerline. Thus, first, in order to update the membership function for the lumen, we calculate the Euclidean distance transform only of the pixels around the extracted centerline. This estimated distance $d_1$ limits the region of interest, since based on its value, we considered only the pixels whose distance transform is lower than $4/p_x$, where $p_x$ is the pixel spacing of the DICOM image and 4 is an approximate value for the radius of the lumen. In other words, we assume that: $f_{lumen} = 0 \rightarrow pixels: d_1 > 4/p_x$ and in a same manner $f_{outer/plaques} = 0 \rightarrow pixels: d_1 > 4/p_x$. For the lumen pixels, the updated membership function is given by:

$$f_1 = f_{lumen} \cdot g^{sigm}(d_1; a_{2,lumen}, b_{2,lumen}), \qquad (7)$$

where $a_{2,lumen} = -0.5$ and $b_{2,lumen} = 2/p_x$, whereas for the outer wall and cp plaques pixels, the updated membership functions are given by:

$$f_{2,outer/2,plaques} = f_{outer/plaque} \cdot g^{sigm}(d_1; a_{2,outer/plaques}, b_{2,outer,plaques}), \qquad (8)$$

where $a_{2,outer/plaques} = -0.5$ and $b_{2,outer/plaques} = 2.5/p_x$.

*2.3.2   Approximation of an initial binary image*

For the implementation of the Level set method approach, an approximation of an initial image-shape $\varphi$ is required. This image is a binary image, which includes 0's as background pixels and 1's as foreground pixels. The intensity threshold value for the estimation of the initial image is $w_i/2$. Thus, the pixel value of initial image is 1, when $f_1 \cdot w_i$ is larger than $w_i/2$, whereas it is 0, when $f_1 \cdot w_i$ is smaller than $w_i/2$. The parameter $w_i$ is an estimated weight to multiply the probability in the level set method and it is defined 1000.

*2.3.3   Calculation of the speed function*

The Level set methods have been widely used in the field of machine vision for segmentation problems, since they are used for the modelling of evolving curves or surfaces. The basic idea behind the level set approaches is the presentation of the interface of a surface, using a higher dimensional function, which is called the level set function. This means that the two dimensional (2D) curve could be described by the 3D level set function, where the additional dimension $t$ represents the time. In the proposed methodology, the detected curve $C \in \Omega$ is represented by the zero level set of a Lipschitz function $\varphi : \Omega \to \mathbb{R}$, such that:

$$\begin{aligned} C &= \partial \omega = \{(x, y, t) \in \Omega : \varphi(x, y, t) = 0\}, \\ inside(C) &= \omega = \{(x, y, t) \in \Omega : \varphi(x, y, t) > 0\}, \\ inside(C) &= \omega = \{(x, y, t) \in \Omega : \varphi(x, y, t) < 0\}, \end{aligned} \qquad (9)$$

where $x, y$ are the spatial coordinates of the 2D image, $t$ is the dimension of time and $\omega \subset \Omega$. In this stage, a level set based variational method using prior shapes is implemented for the lumen segmentation and our aim is to incorporate shape priors into the Chan-Vese's model for segmentation. This approach is based on Cremers *et al.* [26] and Chan *et al.* [25] studies, in which besides the basic level set segmentation function $\varphi$, a shape function $\psi$ and a labelling function $L$ are introduced. The key idea of this methodology is that the defined prior shape is compared with the region where both the level set function $\varphi$ and the labelling function $L$ are positive.

In the presented methodology, the speed function that is impemented to evolve the level set curve is based on Chan *et al.* approach [25]. The defined speed function is a Chan-Vese energy function, combined with prior shapes and with a labelling function and is given by:

$$E(\varphi, \psi, L, c_1, c_2) = E_{cv} + E_{shape} + E_{\psi}, \qquad (10)$$

where $E_{cv}$ is the Chan-Vese energy funtion, $E_{shape}$ is a shape comparison term and $E_{\psi}$ is a labelling term.

The Chan-Vese energy funtion [24] is widely used in medical image segmentation approaches and it is defined as:

$$E_{CV}(c_1,c_2,C) = \int_{inside(C)} (u(x,y)-c_1)^2 dxdy + \int_{outside(C)} (u(x,y)-c_2)^2 dxdy. \quad (11)$$

Disretizing the above speed function equation and writing it as a pixelwise function, it gives

$$E_{CV}(x,y) = (u(x,y)-c_1)^2 - (u(x,y)-c_2)^2, \quad (12)$$

where $u$ is the image, $x, y$ are the spatial coordinates of the 2D image, $C$ is the segmentation curve and $c_1, c_2$ are the average greyscale intensity values inside and outside of $C$, respectively.

The shape comparison term is defined as:

$$E_{shape}(\varphi, L, \psi) = \int_\Omega (H(\varphi)H(L) - H(\psi))^2 dxdy, \quad (13)$$

where $H$ is the Heavyside function and $H(\varphi)H(L)$ represents the intersection of $\varphi > 0$ and $L > 0$. The labelling function $E_\psi$ is a term that indicates if the lumen is segmented successfully, since then this term will be small and it is given by:

$$E_\psi(\psi, c_1, c_2) = \int_\Omega \{(u-c_1)^2 H(\psi) + (u-c_2)^2 (1-H(\psi))\} dxdy. \quad (14)$$

*2.3.4  Sparse Field Algorithm implementation*

As previously mentioned, the key idea of Level Set approaches is that only the area, where $\varphi(x,y) \approx 0$ is important to accurately represent the curve. In this approach, a sparse field algorithm approach, proposed by Whitaker *et al.* [27] is implemented to maintain an accurate and minimal representation of $\varphi$. Once the initial $\varphi$ is defined, the algorithm returns fully initialized arrays for the label map and for an updated $\varphi$. Both arrays are of the same size and the label map records the status of each point. Once the energy function has been computed for the part of the image which is around the centerline, the level sets may be deformed in order to minimize some of the energy function. Thus, a sparse field approach is implemented twice to update $\varphi$ near the zero level set. First, based on the initial $\varphi$ and on a positive factor $a$, which controls the speed and curvature of the level set, the algorithm results in a new $\varphi_{lumen}$. Subsequently, based on the resulted $\varphi_{lumen}$, the algorithm is implemented again, using a higher value of factor *a,* to achieve a smoother lumen shape. In the first step, a lower value of factor *a* was applied, in order to provide fast segmentation in the whole image. In the second implementation of sparse field algorithm, a higher value of factor *a* is selected and applied only at the region of interest in order to provide smooth and accurate segmented objects which depict only the coronary arteries. In the first step, the factor *a* is defined 0.1, whereas in the second step factor *a* is defined 0.6. The value of *a* factor affects only the speed of the segmentation and it does not affect the quality of the segmentation. The number of iterations for the sparse field algorithm impementation is set to 200.

### 2.4 Outer Wall Segmentation

Similarly to the previous step, a Level Set model is implemented for the outer wall segmentation. However, in this stage the initial $\varphi$ is based on the lumen shape, as segmented in the previous stage and on the updated values of outer wall intensities. More specifically, the required $\varphi$ of this stage is also a binary image, which includes 1's for the pixels, where either the $w_i \cdot f_{2,outer} + w_i \cdot f_{2,plaque}$ is higher than $w_i/2$ or the segmented lumen ($\varphi_{lumen}$) has a positive value. Furthermore, the energy function for this stage, is calculated only for the pixels, where $\varphi_{lumen}$ is larger than -0.1. The sparse field algorithm implementation process for the outer wall is exactly the same as in the previous stage described.

### 2.5 Plaque Segmentation

In this stage, the Level Set method is applied in the region of interest (ROI) of the outer wall. The initial $\varphi$ for the plaque segmentation is based only on the updated plaques intensity function. In other words, the initial $\varphi$ is also a binary image, which includes 1's for the pixels, where $w_i \cdot f_{2,plaque}$ is larger than $w_i/2$. For this phase, only a sparse field algorithm implementation is required, as the segmented objects are relatively smaller and the *a* factor is 0.5. An example of the segmentation procedures is demonstrated in Figure 5.

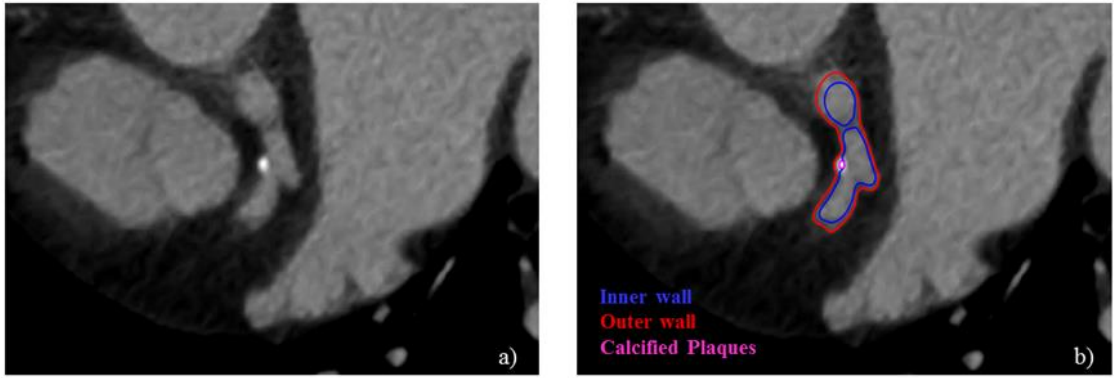

**Figure 5**. An example of the segmentation procedure: a) the acquired image, b) inner wall, outer wall and calcified plaques segmentation.

**2.6. 3D Surface construction**

In this stage, an isosurface of data from each different extracted $\varphi$ array is computed to construct the mesh surfaces. In this step the algorithm marching cubes, proposed by Lorensen and Cline [28], is implemented to construct 3D surfaces for the lumen, outer wall and CP plaques. Marching cubes extracts a polygonal mesh of an isosurface from a 3D discrete scalar field, by proceeding through it. A triangulation approach is implemented, and connecting the detected border points of each CTA image, 3D models are constructed.

**3. Dataset**

The data were acquired from 12 patients, who underwent CTA imaging for clinical purposes and they were used to validate the proposed methodology. Our validation dataset consists of 12 coronary arteries, deriving from six different medical centers. Two arteries were completely healthy (no stenosis present), nine arteries had an intermediate stenosis (seven had a 30%-50% stenosis degree and two had a 50%-70% stenosis degree) and one artery was fully occluded (>90% stenosis degree). Five arteries were scanned with a 64-slice Dual Source Siemens SOMATOM Definition Flash® CT scanner, two arteries were scanned with a Philips Brilliance 64 CT Scanner® and the remaining five arteries were scanned with a 64-slice General Electric Medical Systems Discovery PET/CT 690® scanner.

IVUS and biplane X-ray angiography were acquired from eight patients who underwent coronary catheterization. CTA imaging was also acquired for the same patients. Registration between the imaging modalities was performed using major landmarks common for all modalities, such as the bifurcations or large calcified plaques.

## 4. Artery and CP reconstruction

### 4.1 Artery and CP reconstruction using manual annotations

Reconstruction of arteries was successfully obtained in 12 patients (2 RCA, 8 LAD, 2 LCX) by the proposed methodology in a semi-automatic manner, whereas an expert radiologist manually annotated in the corresponding segments the lumen, the outer wall and two types of plaque (CP plaques and NCP plaques).

The aim of the validation process is to assess the accuracy and the quality of the proposed methodology. In this study, our purpose is to validate our methodology, using as gold standard a medical expert's annotations. It is known that the lack of expert-annotated datasets remains one of the main challenges in medical image processing [29]. Generating high-quality expert-derived annotations in CTA images is time-consuming and requires a specialized in CTA imaging field medical expert.

### 4.2 Artery and CP reconstruction using IVUS

Reconstructions of arteries based on IVUS modality is performed using the study proposed by Bourantas *et al.* [30]. This approach combines the IVUS and X-ray angiography and based on the 3D luminal centerline, derived from two angiographic projections, it places the lumen and the media-adventitial borders detected by IVUS frames onto the centerline.

## 5. Metrics for evaluating 3D image segmentation

The lumen, the outer wall and plaques reconstructed by the two different modalities were compared using as metrics for the 3D image segmentation the Dice Coefficient (DICE) and the Hausdorff Distance (HD) [31]. Additional metrics were used for the validation of the proposed methodology against the IVUS based reconstructed segments. For this purpose we used for comparison two types of Degree of Stenosis (DS1, DS2), the Plaque Burden (PB), the Minimal

Lumen Area (MLA) and the Minimal Lumen Diameter (MLD). More information for all metrics can be found in Appendix.

## 6. Results

For both quantitative evaluation and comparison purposes, we present in Table 2 the Hausdorff distance and the Dice Coefficient distributions obtained by the comparison of the proposed segmentation methodology and the expert's manual segmentation. The HD demonstrates the degree of resemblance between the two models which are superimposed on one another. Thus, the lower HD implies better segmentation, while a higher DICE implies higher accuracy of the segmentation.

**Table 2: Segmentation Validation metrics for CTA images**

| Cases | Arteries | DICE coefficient | Hausdorff Distance |
|---|---|---|---|
| **#1** | LAD | 0.847 | 0.837 |
| **#2** | LAD | 0.675 | 1.910 |
| **#3** | LAD | 0.777 | 0.860 |
| **#4** | LAD | 0.674 | 2.423 |
| **#5** | LAD | 0.759 | 1.589 |
| **#6** | LAD | 0.810 | 1.510 |
| **#7** | LAD | 0.574 | 2.671 |
| **#8** | LAD | 0.751 | 1.800 |
| **#9** | LCX | 0.847 | 1.296 |
| **#10** | LCX | 0.722 | 1.830 |
| **#11** | RCA | 0.781 | 2.341 |
| **#12** | RCA | 0.778 | 1.880 |

The validation process indicates a good agreement, since the mean value of DICE is 0.749, while the mean value of HD is 1.746. In Fig. 6, an example of the lumen and CP 3D models reconstructed by the proposed methodology and by the expert's annotation, is shown. The two reconstructed arteries indicate a similar geometry and plaque distribution. The semi-automated reconstructed artery is smoother than the manually segmented, since a pixel by pixel segmentation may not result in a smooth shape.

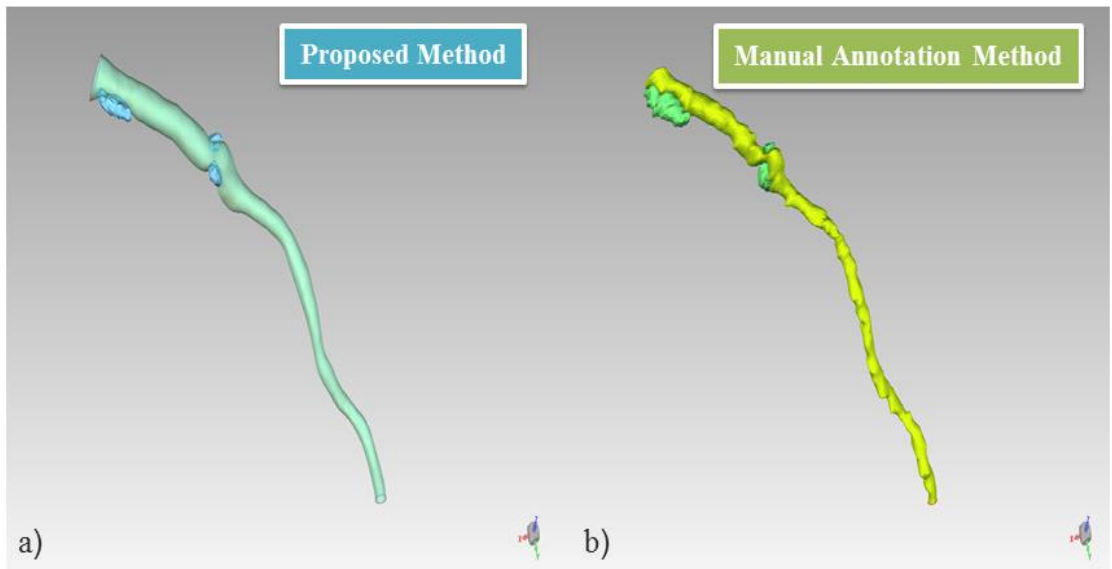

**Figure 6.** Lumen and CP oblects detected by a) the proposed methodology, and b) the medical expert annotation.

As far as the comparison with IVUS process is concerned, we show in Table 3 the values of the comparison metrics. The metrics of the 3D models derived by our methodology are correlated with those derived by the IVUS findings, and the correlation plot for each metric is demonstrated in Fig. 7. It is clear, that the correlation between the two methodologies is statistically significant for all reconstructed cases.

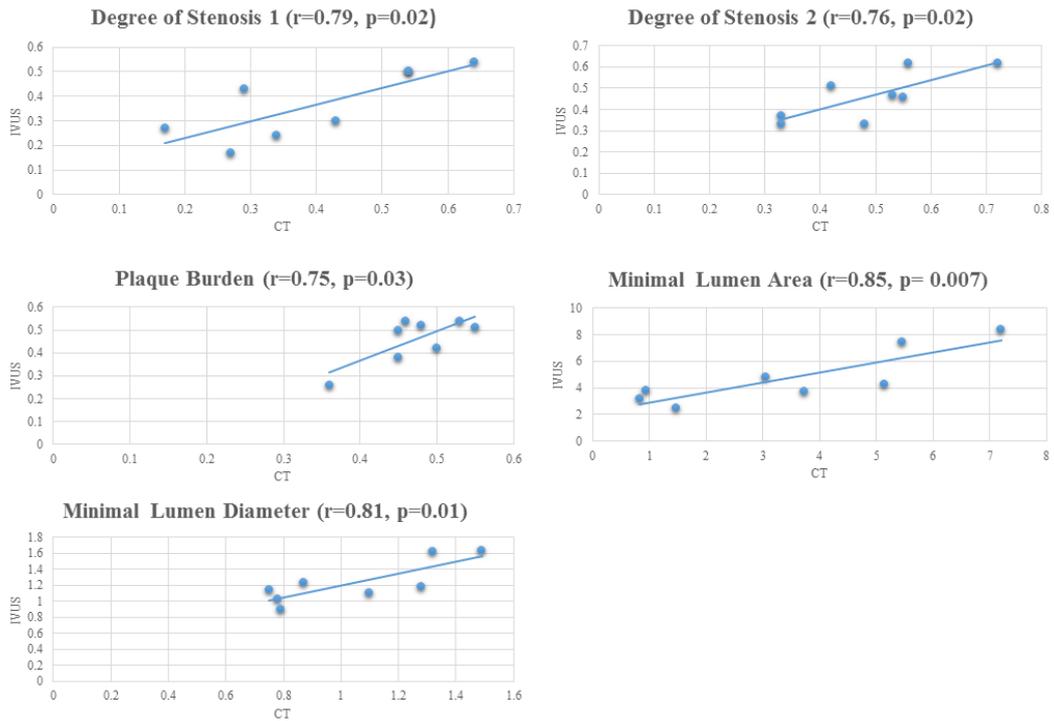

**Figure 7.** Correlation plots for the Degree of Stenosis 1, Degree of Stenosis 2, Plaque Burden, Minimal lumen area, Minimal lumen diameter.

**Table 3: Comparison metrics for CTA images.**

| Cases | Method | DS1 | DS2 | Plaque Burden (/0.5mm) | MLA (mm²) | MLD (mm) |
|---|---|---|---|---|---|---|
| #1 | CT | 0.29 | 0.53 | 0.53±0.09 | 3.73 | 1.10 |
|    | IVUS | 0.43 | 0.47 | 0.54±0.17 | 3.72 | 1.11 |
| #2 | CT | 0.17 | 0.33 | 0.45±0.09 | 7.2 | 1.49 |
|    | IVUS | 0.27 | 0.37 | 0.5±0.07 | 8.37 | 1.64 |
| #3 | CT | 0.64 | 0.55 | 0.45±0.11 | 0.95 | 0.75 |
|    | IVUS | 0.54 | 0.46 | 0.38±0.2 | 3.81 | 1.15 |
| #4 | CT | 0.27 | 0.33 | 0.5±0.07 | 5.46 | 1.32 |
|    | IVUS | 0.17 | 0.33 | 0.42±0.2 | 7.45 | 1.62 |
| #5 | CT | 0.54 | 0.56 | 0.55±0.09 | 0.83 | 0.78 |

| Cases | Method | DS1 | DS2 | Plaque Burden (/0.5mm) | MLA (mm$^2$) | MLD (mm) |
|---|---|---|---|---|---|---|
|    | IVUS | 0.5  | 0.62 | 0.51±0.08 | 3.18 | 1.03 |
| #6 | CT   | 0.34 | 0.48 | 0.46±0.09 | 5.14 | 1.28 |
|    | IVUS | 0.24 | 0.33 | 0.54±0.09 | 4.28 | 1.19 |
| #7 | CT   | 0.54 | 0.72 | 0.48±0.08 | 3.05 | 0.87 |
|    | IVUS | 0.5  | 0.62 | 0.52±0.09 | 4.79 | 1.24 |
| #8 | CT   | 0.43 | 0.42 | 0.36±0.09 | 1.47 | 0.79 |
|    | IVUS | 0.3  | 0.51 | 0.26±0.22 | 2.51 | 0.9  |

## 7. Implementation

The CT segmentation methodology was initially implemented in Matlab R2016a. This allows to re-run all segmentation results and create new models with method improvement. A framework for batch processing all vessels has been developed using a set of inputs (vessel seed points). An experienced user needs less than 30 seconds for the manual component selection, whereas the time complexity of the proposed methodology is about 30 seconds for an arterial segment of 40 mm length using a desktop computer with an Intel core i7 processor and 16 GB of RAM.

## 8. Discussion

In this work, a semi-automated approach for the reconstruction of the lumen, outer wall and CP plaques in coronary arteries, using CTA images, was presented. The main innovation of this study is its semi-automatic nature, since it only requires the starting and the ending point of the coronary artery to accurately reconstruct the artery. The presented approach is mainly based on a level set segmentation model [25] for the segmentation of the lumen, outer wall and CP plaques, contrary to Athanasiou *et al*. study in which the lumen, outer wall and calcified plaque models are reconstructed, based on a classification of the Region of Interest scheme. To our knowledge, our study is the only approach in the literature that allows a 3D reconstruction of coronary anatomy and plaque characterization, which is compared both by a medical expert's annotations and IVUS findings. The results of the proposed methodology demonstrated that our approach is able to accurately segment the lumen, outer wall and provide a reliable detection of CP plaques and geometrically correct 3D models.

In the literature, several methodologies [6, 15, 32-34] have been presented for the segmentation of CTA cross sectional images and the classification of its plaque components. These approaches are time consuming, since corrections of the detected borders are sometimes required and the reconstruction of the coronary anatomy is not implemented in an automatic or semi-automatic manner. This limitation is overcome by the new proposed semi-automatic methodology, which provides in a fast manner the segmentation of CTA images and the detection of CP plaques.

Furthermore, existing studies [15, 16, 18], which are evaluated using only Intravascular Ultrasound (IVUS) modality, whereas the presented study is validated using both medical expert annotation and IVUS 3D models. Although, the manual annotation requires a lot of effort and time, since the expert annotates in each slice the lumen, the outer wall and the plaques, expert manual segmentation of real images is regarded as a practical gold standard against which new segmentation algorithms can be compared [29]. On the other hand, a comparison with IVUS 3D reconstructed models allow us to validate the geometry of the reconstructed arteries and extract validation metrics, such as the coronary lumen stenosis, the plaque burden, the minimal lumen area and the minimal lumen diameter.

Additionally, in contrast to Athanasiou *et al.* existing study [18], which focused on comparison metrics, such as the comparison of the volume and areas of the ROI, the length and angle of the vessel, our validation process is dedicated to quantify the region of agreement, the overlap region between our proposed methodology and the manual segmentation, as well as validation metrics which demonstrate the 3D model accuracy. It is known that an objective validation of image segmentation is of great importance, but it is such a difficult task. Based on the literature [29, 31], the selected in this approach comparison metrics are two of the most common measures and their values indicate how accurately the segmentation algorithm performs. More specifically, the mean value of DICE is 0.749, while its standard deviation is 0.0787 and the mean value of HD is 1.746, while its standard deviation is 0.573. The correlation coefficients for the DS1, the DS2, the plaque burden, the MLA and MLD, when comparing the derived from the proposed methodology 3D models with the IVUS reconstructed models, were 0.79, 0.77, 0.75, 0.85, 0.81, respectively

In the proposed methodology, the validation dataset consists of data acquired from three different scanners. That means that our 3D reconstruction approach is applicable in different clinical environments, since each Computed Tomography (CT) scanner is characterized by its different properties and settings.

The presented Level Set-based segmentation methodology allows an accurate segmentation and its applicability is not limited by the low CTA images quality [24]. The unique preprocessing step of the acquired CTA images is to detect the vessel candidate regions. In addition to this, the extraction of the centerline using a minimum cost path based approach in combination with appropriate cost functions selection, ensures that the extracted centerline may be the globally optimal solution. Therefore, once the vessel centerline is successfully extracted, an appropriate initial vessel mask for the lumen segmentation is created. Furthermore, minimum cost path approaches are able to overcome problems related to overlapping pathologies depicted in the image, as well as, issues of low image quality.

Although the presented methodology provides an accurate 3D reconstruction of coronary arteries and detection of CP plaques, there is a need to improve and further validate the algorithm performance. More specifically, the proposed methodology does not take the blooming artifact into consideration. The main challenge of the clinician is to quantify the vessel reduction, when CP plaque is present. However, the densely calcified plaques create a blooming effect on CTA images, which limits the ability to accurately identify the lumen contours and sometimes the real degree of stenosis is overestimated, resulting in an unnecessary use of invasive coronary angiography. Thus, the incorporation of blooming artifact removal is one of our future steps to enhance our diagnostic confidence in the term of vessel stenosis.

Furthermore, it is observed that the discrimination of the lumen, the outer wall and the CP plaques requires different parameters definition. In this study, these parameters have been defined heuristically. One of our most basic future steps is the adaptation of all these parameters into the CTA acquired image. In other words, our future goal is to define an automatic parameter tuning procedure for the discrimination of the lumen, outer wall and CP plaques. Thus, it would be easier for the user to achieve an accurate vessel segmentation and CP plaques detection and since the parameters would be ideally adapted to the initial CTA image intensities, a higher accuracy might be observed.

## 9. Conclusions

The presented methodology for 3D reconstruction of coronary arteries and atherosclerotic plaques achieves an expetide coronary reconstruction and a detailed coronary anatomy and plaque distribution representation on real CTA images. The main innovation of the proposed methodology is its semi-automated nature, since it requires the minimal user interaction to reconstruct accurately the coronary arteries. The validation procedure indicates that the presented methodology accurately constructs the 3D models of lumen, outer wall and CP plaques. The incorporation of future refinements tasks, such as the blooming effect removal and the adaptation of algorithm's parameters into the different CTA images, may provide useful real time analysis of coronary arteries and a wide clinical applicability.

## Acknowledgements

This work is partially funded by the European Commission: Project SMARTOOL, "Simulation Modeling of coronary ARTery disease: a tool for clinical decision support — SMARTool" GA number: 689068).

**Appendix**

   **1. Metrics for the validation**

DICE, also called overlap index, is an overlap based metric. DICE is widely used to compare directly automatic and expert's annotations segmentations. It is considered as a statistical validation metric to evaluate the performance of both the reproducibility of manual segmentations and the spatial overlap accuracy of automated segmentation [35]. DICE is defined as:

$$DICE = \frac{2\left|S_g^1 \cap S_t^1\right|}{\left|S_g^1\right| + \left|S_t^1\right|},$$

where $S_t^1$ presents the automatic segmentation and $S_g^1$ presents the expert's annotations.

HD is a spatial distance based metric and is commonly used in the evaluation of image segmentation as dissimilarity measure. HD between two finite point sets A and B is defined as:

$$HD(A,B) = \max(h(A,B), h(B,A)),$$

where $h(A,B) = \max_{a \in A} \min_{b \in B} \|a - b\|$.

The 3D models reconstructed by CTA and IVUS were compared using as validation metrics two types of Degree of Stenosis (DS1, DS2), the Plaque Burden (PB), the Minimal Lumen Area (MLA) and the Minimal Lumen Diameter (MLD). Each 3D model was sliced every 0.5 mm, to estimate the validation metrics. Based on the literature, two different ways of Degree of Stenosis estimation were used [36]. More specifically, Degree of Stenosis1 (DS1) and Degree of Stenosis 2 (DS2) is given by:

$$DS1 = \frac{B - A}{B} \cdot 100\%,$$

$$DS2 = \frac{C - A}{C} \cdot 100\%,$$

where $A$ is the luminal diameter at the site of maximal narrowing, $B$ is the diameter of the normal distal coronary artery beyond the bulb where the artery walls are intersected and $C$ is the diameter of estimated original width of the coronary artery at the site of maximal narrowing. The Plaque Burden (PB) is extracted based both on the inner and the outer wall contours areas per 0.5 mm and is given by:

$$PB = \frac{outer\_wall\_area - inner\_wall\_area}{outer\_wall\_area}.$$

2. **Threshold selection**

The selection of the intensity thresholds of the lumen and the calcified plaques were based on the current literature. More specifically, due to the variety of CT scanners, the average range of lumen intensity is 200 HU–500 HU[1], whereas the intensity of calcified plaques is higher than 500 HU[1]. In this methodology, we have to identify the optimal threshold values for the discrimination of the lumen, the outer wall and the calcified plaques. These threshold values are extracted based on the calculation of the mean lumen intensity ($I_{lumen}$), in combination with the known ranges of the lumen and calcified plaques intensities. Thus, as far as the lumen threshold is concerned and considering that the intensity of the lumen is affected by the acquisition dose protocol, we selected a relative small $l_{thres}$ ($l_{thres}$ =80HU). As it is demonstrated in Table 1 (page 10), in order to find the lower limit of the lumen intensity value, the $l_{thres}$ is subtracted from the mean lumen intensity and the value of 80 HU is a good approximation in order to agree with the lumen range proposed from the literature (around 500HU). As it is demonstrated in the Figure 4, the HU values which are possible to match with the lumen are about 500 HU. Furthermore, several experiments have been implemented using $l_{thres}$ values close to 80HU in order to examine the algorithm effectiveness for the whole range of the membership function demonstrated in Figure 4 (page 10). A similar approach has been implemented for the calcified plaques. More specifically, according to current literature HU>500 may identify calcified plaques. For this purpose our approach was implemented after the detection of the lumen border and especially at the region out of the lumen border. In a similar approach the $cp_{thres}$ should be 400HU for optimal accuracy in calcified plaque detection.

As far as the attenuation HU value for non calcified plaques is concerned, it depends also on the contrast protocol. However, based on the literature a potential range for non-calcified plaques is from 0 to100 HU[3-9]. Thus, the threshold value of non-calcified plaques is defined to be 50 HU, in order to include HU values around 50 HU, depending on the density of the lumen. This selected threshold value is considered as an indicative value for the non-calcified plaques, which is adapted to the lumen density.